\documentclass[fleqn,twoside, epsf]{article}
\usepackage{espcrc2}

\usepackage{epsfig}
\usepackage{graphics}
\usepackage[figuresright]{rotating}

\newcommand{\AmS}{{\protect\the\textfont2
  A\kern-.1667em\lower.5ex\hbox{M}\kern-.125emS}}
\title{
Glueball Matrix Elements on
Anisotropic Lattices\thanks{Presented by S.-J. Dong.}}
\author{Y. Chen\address[UK]{Department of Physics and Astronomy, 
University of Kentucky, Lexington, KY 40506, USA}\address{Institute of High Energy Physics, Chinese Academy of 
Sciences, Beijing 100039, P. R. China},
        S.-J. Dong\addressmark[UK],
        T. Draper\addressmark[UK],
	I. Horv\'{a}th\addressmark[UK],
	F.-X. Lee\address{Center for Nuclear Studies, Dept. of Physics, 
George Washington Univ., Washington, DC 20052, USA}\address{Jefferson 
Lab, 12000 Jefferson Avenue, Newport News, VA 23606, USA},
	N. Mathur\addressmark[UK],
	C. Morningstar\address{Department of Physics, Carnegie Mellon 
University, Pittsburgh, PA 15213, USA},\\
	M. Peardon\address{School of Mathematics, Trinity College, 
Dublin 2, Ireland},
	S. Tamhankar\addressmark[UK],
	B.L. Young\address{Department of Physics, Carnegie Mellon 
University, Pittsburgh, PA 15213, USA}, and
	J.-B. Zhang\address{CSSM and Department of Physics, Univ. of 
        Adelaide, Adelaide, SA 5005, Australia}
         }
       
\begin{document}

\begin{abstract}
The glueball-to-vacuum matrix elements of local gluonic 
operators in scalar, tensor, and pseudoscalar channels are investigated 
numerically on several anisotropic lattices with the spatial lattice 
spacing in the range 0.1$fm$ -- 0.2$fm$. These matrix elements are needed 
to 
predict the glueball branching ratios in $J/\psi$ radiative decays which 
will help to identify the glueball states in experiments.
Two types of improved local gluonic operators are
constructed for a self-consistent check, and the finite volume 
effects are also studied. The lattice spacing dependence of our results is
very small and the continuum limits are reliably extrapolated.
\vspace{1pc}
\end{abstract}

\maketitle

\section{Introduction}

Glueballs, predicted by QCD, are so exotic from the point of view of
the naive quark model that their existence will be a confirmation of QCD. 
Extensive numerical studies have been carried out to simulate the glueball 
spectrum and resulted that the low-lying
glueballs are in the mass range 1 -- 3 GeV, which suggests that the 
$J/\psi$ 
radiative decays are the ideal hunting ground for glueballs. There are 
several possible glueball candidates in the final states of $J/\psi$ 
radiative decays, however, more criteria are needed for their 
unambiguous identifications, one of which might be the partial widths of 
$J/\psi$ radiative decays into glueballs. To estimate these partial 
widths, the vacuum-to-glueball transition matrix elements(TME) of local 
gluonic operators should be derived first.
\par
The techniques of lattice simulations in the glueball sector have been
substantially improved in the past decade. Inspired by the success of 
anisotropic lattice techniques in the simulations of the glueball 
spectrum~\cite{spectra} and as a continuation of former 
studies~\cite{matrix,dong}, this work is devoted to the numerical study of 
TME on anisotropic lattices with tadpole-improved gauge action.

\section{LOCAL GLUONIC OPERATORS}

The TME computed in this work are $\langle 0|S(x)|0^{++}\rangle$, 
$\langle 0|T_{\mu\nu}(x)|2^{++}\rangle$, and $\langle 
0|P(x)|0^{-+}\rangle$, where $|J^{PC}\rangle$ refers to the glueball state 
with the quantum number $J^{PC}$, and the local operators $S(x)$, 
$T_{\mu\nu}$, and $P(x)$ are trace anomaly $g^2 Tr 
G_{\mu\nu}G_{\mu\nu}(x)$, the energy-momentum tensor $  
g^2 Tr(G_{\mu\alpha}(x)G_{\alpha\nu}(x)
-\frac{1}{4}g_{\mu\nu}G^2(x))$, and the topological charge density 
$ g^2 \epsilon_{\mu\nu\rho\sigma}Tr
G_{\mu\nu}(x)G_{\rho\sigma}(x)$, respectively ($G_{\mu\nu}(x)$ is the 
gauge field strength and $g$ the gauge coupling). 
Two types of lattice local gluonic operators (Type-I and Type-II) are 
constructed in this work. 
\par
Type-I operators are linear combinations of a set of small Wilson loops 
according to the irreducible representations of the lattice symmetry 
group, namely, $A_1^{++}$, $A_1^{-+}$, $E^{++}$, and 
$T_2^{++}$~\cite{berg}. Letting  
$O^{R}(x)$ be the local operators of specific quantum number $R$, 
the construction of Type-I operators can be expressed as
\begin{equation}
O^{R}(x) = \sum\limits_{i} C_i^{R} ReTr(W_1^{(i)}(x) + \alpha 
W_2^{(i)}(x) + \ldots ),
\end{equation}  
where $C_i^{R}$ are the combinational coefficients. Different Wilson loops
$W_1, W_2,\ldots$ are included with proper factors $\alpha$ to improve the
operator. 
\par
To construct Type-II operators, we define the lattice gauge field strength
$\hat{F}_{\mu\nu}(x)$ as
\begin{equation}
\hat{F}_{\mu\nu}(x) =Im\langle f(u_s)P_{\mu\nu}(x) + g(u_s) 
R_{\mu\nu}(x)\rangle_c,
\end{equation}
where $P_{\mu\nu}(x)$, and $R_{\mu\nu}(x)$ are respectively the plaquette 
and rectangle at $x$. $\langle |\rangle_c$ means the clover average, 
$f(u_s)$ and $g(u_s)$ are factors including the 
tadpole parameter $u_s= (\langle 1/3 Tr P_{ij}\rangle)^{1/4}$ and are 
chosen in a way that yields $\hat{F}_{\mu\nu}(x)=a_s^2 (G_{\mu\nu}(x)+ 
O(a_s^4))$ ($a_s$ is the spatial lattice spacing). Proper combinations
of $Tr\hat{F}_{\mu\nu}(x)\hat{F}_{\rho\sigma}(x)$ give the Type-II 
operators with reduced lattice artifacts.
\begin{table}[t]
\caption{The simulation parameters.}
\label{table:1}
\setlength{\tabcolsep}{1.5mm}
\begin{tabular}{ccccrr}
\hline
$\beta$ & $\xi$ & $u_s$ & $a_s(fm)$ & $L^3\times T$ & \#Meas. \\\hline
   2.4  &   5   & 0.409 & 0.222(2)  & $8^3\times 40$ & 20000 \\
        &       &       &           &$12^3\times 64$ & 10000 \\
        &       &       &           &$16^3\times 80$ & 10000 \\
   2.6  &   5   & 0.438 & 0.176(1)  &$12^3\times 64$ &  8600 \\
   2.7  &   5   & 0.451 & 0.156(1)  &$12^3\times 64$ & 10000 \\
   3.0  &   3   & 0.500 & 0.120(1)  &$16^3\times 48$ & 10000 \\
   3.2  &   3   & 0.521 & 0.101(1)  &$24^3\times 72$ &  7900 \\
\hline
\end{tabular}
\end{table}
\par There are seven matrix elements calculated in this work, which are
denoted by (S,B), (S,E), (E,B), (E,E), (T,B), (T,E), and (PS). Here ( ,B) 
(or ( ,E)) means the operator is made up of color-magnetic (or 
color-electric) field, and (PS) refers to the pseudoscalar channel.
\begin{figure}[t]
\begin{center}
\epsfxsize=2.5in\epsfbox[80 100 530 490]{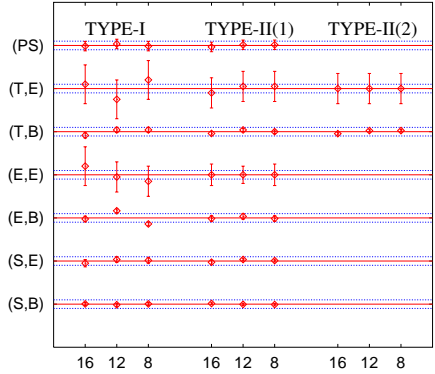}
\end{center}
\vspace{-0.5cm}
\caption{Finite-volume effects of matrix elements.}
\label{fig1}
\end{figure}

\section{SIMULATION DETAILS AND RESULTS}  
Numerical simulations were carried out on anisotropic 
lattices with tadpole-improved gauge action~\cite{spectra}. Five 
independent 
simulations have been done with input parameter listed in Table. 1. 
\par
With the tadpole improvement, the local gluonic operators on the lattice 
are all improved to $O(a_s^4)$ ($a_s$ is the spatial lattice spacing) at 
the 
tree level. With applying the variational method to the combinations of 
smeared Wilson loops with different prototypes~\cite{spectra}, glueball 
states are obtained through correlators of smeared operators which have 
large overlaps with glueball states. Six independent runs were 
carried 
out, on lattices with spatial lattice spacings in the range 0.1 $fm$ -- 
0.22 $fm$, to measure the smeared-smeared correlators  $C_{SS}(t)$ and 
smeared-local 
correlators $C_{SL}(t)$. The matrix elements are extracted by 
fitting $C_{SS}(t)$ and $C_{SL}(t)$ simultaneously using the correlated 
$\chi^2$ method. The fit models are taken as 
\begin{eqnarray}
C_{SS}(t)&=& X^2 e^{-Mt}\\\nonumber
C_{SL}(t)&=& XY e^{-Mt},
\end{eqnarray}
where $X$ is the amplitude of the glueball operators,
$Y$ is the glueball-to-vacuum matrix element, and $M$ is the
glueball mass. 
\par
The finite volume effects (FVE) of matrix elements are studied on 
lattices 
$8^3\times 40$, $12^3\times 64$, and $16^3\times 80$ at $\beta=2.4$, 
$\xi=5$, and the results are shown in Fig.\ \ref{fig1}, where each 
point with a errorbar is the fractional change, $\delta_G(L)
=1-f_G(L)/\bar{f}_G$, in the matrix elements ($\bar{f}_G$ is the 
average value of the matrix elements of glueball $G$ over the three lattice 
volumes, and $f_G(L)$ is the matrix element of glueball $G$ measured on 
lattice $L^3\times T$). The labels $L=8, 12$, and $16$ denote the 
different lattice volumes, and the labels along the vertical axis 
represent the matrix elements of the different local operators. 
To guide eyes, $\delta_G = 0$ and $\delta_G = \pm 0.02$ are also drawn in
Fig.\ \ref{fig1} with solid line and dash lines, respectively.  
All changes are statistically consistent with zero, indicating that 
systematic errors from FVE are negligible.
\begin{figure}[t]
\begin{center}
\epsfxsize=2.3in\epsfbox[80 100 530 490]{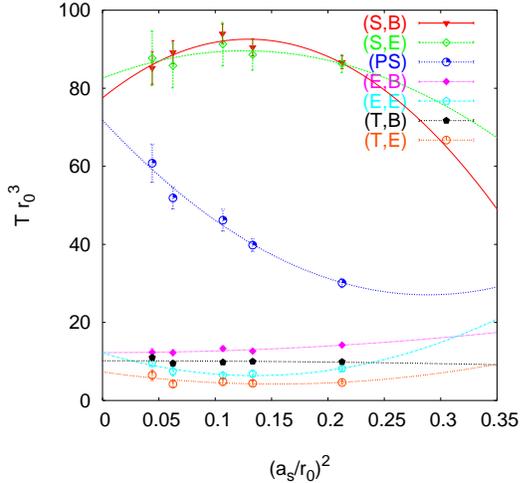}
\end{center}
\vspace{-0.5 cm}
\caption{
Continuum extrapolations of matrix elements of Type-I 
operators.
}
\label{fig2}
\end{figure}
\begin{figure}[t]
\begin{center}
\epsfxsize=2.3in\epsfbox[80 100 530 490]{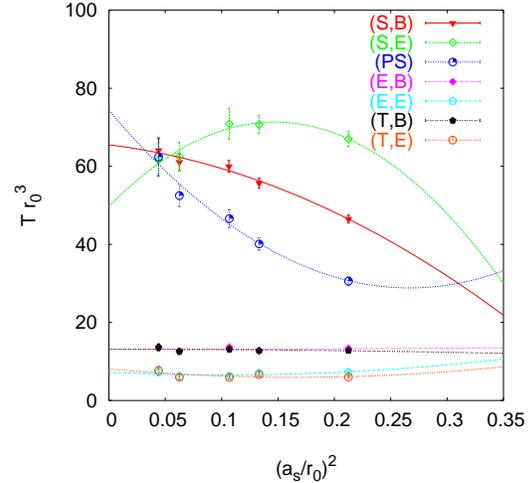}
\end{center}
\vspace{-0.5 cm}
\caption{
Continuum extrapolations of matrix elements of Type-II 
operators.
}
\label{fig3}
\end{figure}
\par The simulated results of matrix elements at different lattice 
spacings 
are shown in Fig.\ \ref{fig2} (Type-I operators) and Fig.\ 
\ref{fig3} (Type-II operators), where $a_s$ (in units of $r_0\sim 
(410(20)MeV)^{-1}$, the hadronic scale parameter) dependences are 
observed.  The matrix elements can be reliably extrapolated to 
the continuum limit by using the form
$f(a_s/r_0) = Tr_0^3 + c_2 (a_s/r_0)^2 +c_4 (a_s/r_0)^4$,
where $Tr_0^3$, $c_2$, and $c_4$ are best-fit parameters. We keep the 
$a_s^2 $ term in the fitting model because there are residual 
$\alpha_s a_s^2$ 
artifacts in the gauge action and some local operators. From the figures 
one can find that the Type-II operators exhibit better behaviors (for 
example, the matrix elements of $T_2$ and $E$ representations 
coincide, as it
should be when the rotational symmetry is restored in the continuum 
limit).
\par The physically available predictions will not be derived until the 
local gluonic operators are properly renormalized. The nonperturbative
renormalization of these operators is in progress. 
\par
This work is supported by DOE Grants DE-FG05-84ER40154 and
DE-FG02-02ER45967. Y. Chen is partly 
supported by NSFC (No.10075051, 12035040) and CAS (KJCX2-SW-N02).

\end{document}